\documentstyle[prb,aps]{revtex}

\begin{document}

\draft
\twocolumn
[
\title{Field-induced incommensurate-to-commensurate transition in
Ba$_2$CuGe$_2$O$_7$.}

\author{A. Zheludev \and S. Maslov \and G. Shirane}

\address{Brookhaven National  Laboratory,
Upton, NY 11973-5000, USA.}

\author{Y. Sasago, N. Koide, and K. Uchinokura}

\address{Department of Applied Physics, The University of Tokyo,\\
 7-3-1 Hongo, Bunkyo-ku, Tokyo 113, Japan.}

\date{\today}

\maketitle
\widetext
\advance\leftskip by 57pt

\begin{abstract}

We report an observation of a commensurate-incommensurate phase
transition in the Dzyaloshinskii-Moriya spiral antiferromagnet
Ba$_2$CuGe$_2$O$_7$. The transition is induced by an external  magnetic
field applied along the $c$ axis of the tetragonal structure, i. e., in
the plane of spin rotation. Bulk magnetic measurements and neutron
diffraction experiments show that the transition occurs in a critical
field $H_{c}\approx 2.1$~T. Experimental results for the period of the
magnetic structure and magnetization as functions of magnetic field are
in quantitative agreement with our exact analytical solution for
Dzyaloshinskii's model of commensurate-incommensurate transitions in
spiral magnets.

\end{abstract}
]
\narrowtext

\section{introduction}

Among relatively unusual and exotic magnetic interactions in solids is
the Dzyaloshinskii-Moriya (D-M) asymmetric exchange
interaction.\cite{Dzya58,Dzya64,Moriya60} Unlike the conventional
Heisenberg exchange coupling, it is proportional to the {\it vector}
product of interacting spins, and is permitted by symmetry only in
non-centric crystal structures. It results from relativistic spin-orbit
corrections \cite{Moriya60} to the ordinary superexchange mechanism
\cite{Anderson59} and therefore is usually weak compared to
antiferromagnetic symmetric exchange. This is why only few materials in
which D-M interactions play an important role have been found so far.
The best known examples are cubic FeGe\cite{Wilkinson76,Lebech89} and
MnSi,\cite{Ishikawa76,Ishikawa77,Ishikawa84} where D-M terms in the
Hamiltonian cause an instability of ferromagnetic order towards the
formation of an incommensurate spiral structure.

This paper deals with the properties of Ba$_2$CuGe$_2$O$_7$, a newly
investigated system in which D-M coupling plays a key role. Originally
the magnetic properties of this quasi 2-dimensional (2D) antiferromagnet
(AF) were investigated as part of the ongoing search for new singlet
ground state compounds, triggered by the discovery of a spin-Peierls
transition and other extraordinary magnetic properties of
CuGeO$_{3}$.\cite{Hase93L,Hase93B} While some new materials related to
CuGeO$_{3}$, e.g., CaCuGe$_2$O$_{6}$\cite{Sasago95,Zheludev96-CACUGEO}
and BaCuSi$_{2}$O$_{6}$,\cite{Sasago96-BACUSIO} have dimerized ground
states and energy gaps in their spin excitation spectra,
Ba$_2$CuGe$_2$O$_7$ undergoes a transition to a magnetically ordered
phase below $T_{N}\approx3.2$~K and the magnetic excitations are gapless
spin waves.\cite{Zheludev96-BACUGEO} Even though the magnetic properties
of Ba$_2$CuGe$_2$O$_7$ can be adequately described in terms of classical
spins, they are rather intriguing: the magnetic structure is an
incommensurate spin-spiral. By performing detailed measurements of the
spin-wave dispersion we have previously demonstrated that this
spiral ordering may not be caused by competing exchange
interactions.\cite{Zheludev96-BACUGEO} Since Ba$_2$CuGe$_2$O$_7$ has a
non-centric crystal structure, we have suggested that the incommensurate
magnetic phase is a result of Dzyaloshinskii-Moriya interactions.

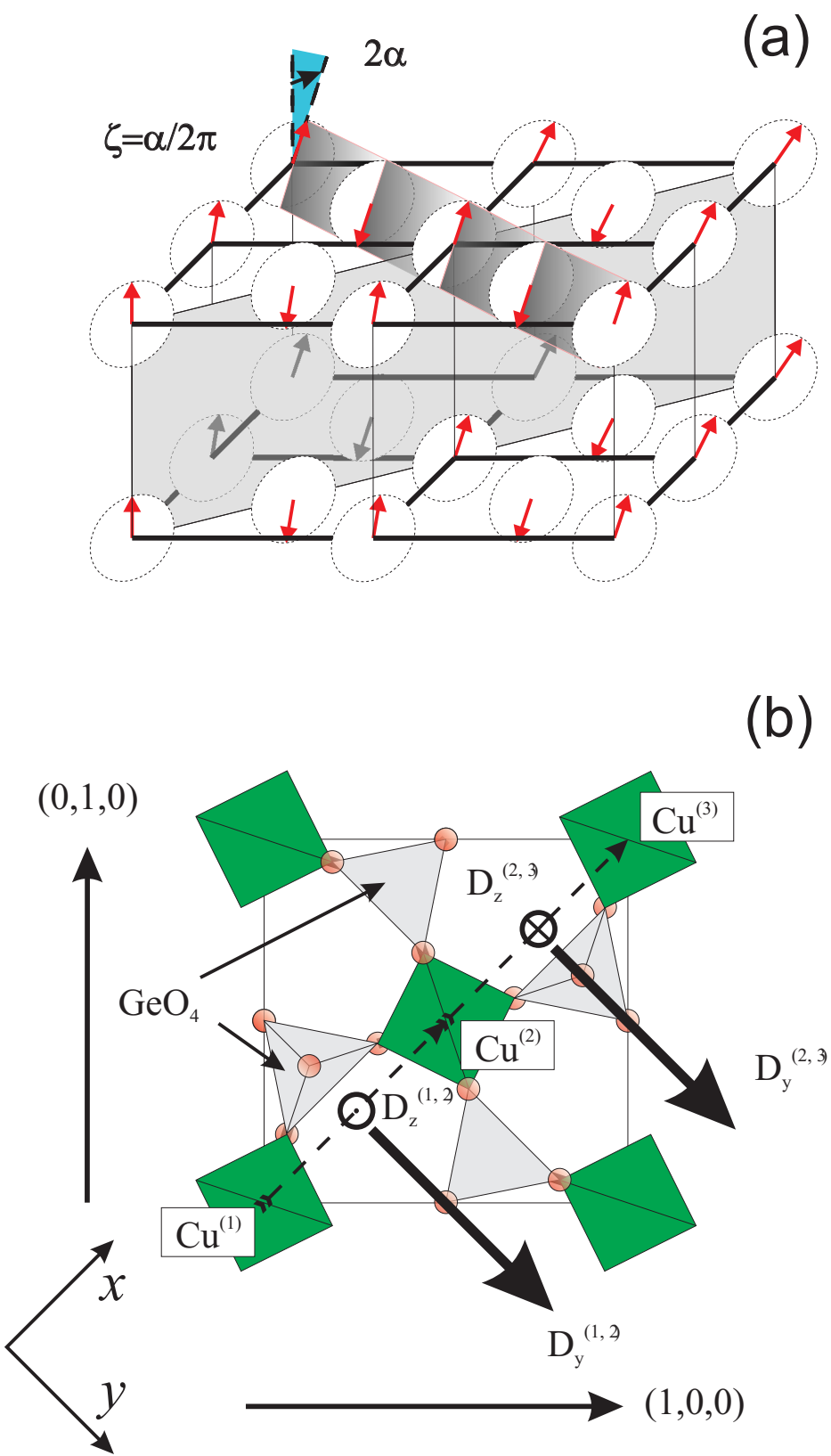
\begin{figure}
\caption{a) Magnetic structure of Ba$_{2}$CuGe$_{2}$O$_{7}$
(Ref.~\protect\onlinecite{Zheludev96-BACUGEO}). b)  D-M interactions in the
Cu-planes of Ba$_{2}$CuGe$_{2}$O$_{7}$. The (1,-1,0)-component of the
Dzyaloshinskii vector ${\bf D}$ (solid arrows) is the same
for all oriented Cu-Cu bond (dashed arrow) along the $(1,1,0)$
direction. The $z$-component is sign-alternating.}
\label{DM}
\end{figure}

The crystal structure of Ba$_2$CuGe$_2$O$_7$, as well as the spiral spin
arrangement in the magnetically ordered phase, were discussed in detail
in our previous work,\cite{Zheludev96-BACUGEO} and only the essential
features are reviewed here. The magnetic ($S=1/2$) Cu$^{2+}$ ions are
arranged on a square lattice in the $(a,b)$ plane of the tetragonal
structure [space group P$\overline{4}2_1{\rm m}$ (No.~113), lattice
constants $a=8.466$~\AA, $c=5.445$~\AA]. The in-plane interactions
between spins are established through GeO$_{4}$ tetrahedra that,
together with the Cu$^{2+}$ ions, form distinct Cu-Ge-O layers. These
layers are well isolated by interstitial non-magnetic planes of
Ba$^{2+}$ ions. Only nearest-neighbor in-plane antiferromagnetic
exchange interactions are important ($J\approx0.48$~meV). Interplane
coupling is ferromagnetic and substantially weaker
($J_{\bot}\approx0.013$~meV). In the ordered phase the spins lie in the
$(1,\overline{1},0)$ plane and the propagation vector for the spiral is
$(1+\zeta,\zeta,0)$ where $\zeta=0.027$ [Fig.~\ref{DM}(a)]. The magnetic
structure is a distortion of a N\'{e}el spin arrangement: a translation
along the $(1,1,0)$ direction (hereafter referred to as the $x$ axis)
induces a rotation of the spins by an angle
$\alpha=\case{2\pi}{\zeta}\approx9.7^{\circ}$ in the
$(1,\overline{1},0)$ plane (relative to an exact antiparallel
alignment). Along the $(1,\overline{1},0)$ direction ($y$-axis) the
spins are perfectly antiparallel. Nearest-neighbor spins from adjacent
Cu-planes are aligned parallel to each other.

The mechanism by which D-M interactions can stabilize the spiral structure
in
Ba$_2$CuGe$_2$O$_7$ is illustrated in Fig.~\ref{DM}(b). The D-M energy for
two interacting spins ${\bf S}_{1}$ and ${\bf S}_{2}$ may be written as
$({\bf S}_{1}\times {\bf S}_{2})\cdot {\bf D}^{(1,2)}$, where ${\bf
D}^{(1,2)}$ is the so-called Dzyaloshinskii vector attributed to the oriented
bond between the two spins and ``$\times$'' denotes a vector product. For
spins lying in the $(1,\overline{1},0)$ plane the only relevant component
of the  vector ${\bf D}$ is $D_{y}$. The symmetry of the structure is such
that for two subsequent Cu-pairs $(1,2)$ and $(2,3)$ along the $(1,1,0)$
direction $D^{(1,2)}_{y}=D^{(2,3)}_{y}$ [Fig.~\ref{DM}(b)]. A finite value
for D energetically favors a $90^{\circ}$ angle between subsequent spins.
Since AF exchange favors an angle of $180^{\circ}$, the total energy is
minimized at some intermediate angle that is defined by the ratio of
exchange and D-M interactions strengths.

How can the mechanism described above be verified experimentally? If one
could force the spins into the $(0,0,1)$ plane for example, $D_{y}$
would become inactive, due to the nature of the vector product, and
the incommensurate spiral would disappear. The
$z$-component of ${\bf D}$, i.e., the $(0,0,1)$-projection, on the
other hand, would become
relevant. As shown in Fig.~\ref{DM}(b), for the $D_{z}$, unlike for
$D_{y}$, symmetry dictates a {\it change of sign} from one Cu-Cu bond to
the next: $D^{(1,2)}_{z}=-D^{(2,3)}_{z}$. As a result, D-M interactions
would tend to distort the N\'{e}el spin arrangement towards a
weak-ferromagnet (canted) structure, that is {\it commensurate} with the
crystal lattice. In practice forcing spins into the $(a,b)$ plane can be
achieved by applying an external magnetic field along the $c$-axis of
the crystal. Indeed, in  structure with zero net magnetic moment
spins tend to align
perpendicular to the external field, which in the simplest case leads to
spin-flop transitions in conventional antiferromagnets. In other words,
if the proposed model for D-M interactions in Ba$_2$CuGe$_2$O$_7$ is
correct, we expect a field-induced commensurate-incommensurate magnetic
transition (CI) in this material.

In the present paper we report an experimental observation of such a
phase transition in Ba$_2$CuGe$_2$O$_7$ by means of bulk magnetization
and neutron diffraction measurements. We find that the phase transition
is of rather unusual character and is the first ``clean'' realization of
Dzyaloshinskii's model for CI transitions in spiral magnets induced by a
magnetic field applied {\it in the plane of spin rotation}.\cite{Dzya65}
We also present an exact continuous-limit solution to Dzyaloshinskii's
model. For the particular case of Ba$_2$CuGe$_2$O$_7$ theoretical
predictions are in quantitative agreement with experimental data. A
short preliminary report on this work is published
elsewhere.\cite{prelim}

\section{Experimental procedures}
Transparent, slightly yellowish single-crystal samples were prepared
using the floating-zone method. Magnetization measurements were
performed with a conventional DC-squid magnetometer in the temperature
range 2--300~K. Two sets of neutron scattering experiments were carried
out on the H4M (thermal beam) and H9 (cold beam) 3-axis spectrometers at
the High Flux Beam Reactor (HFBR) at Brookhaven National Laboratory on
an irregularly shaped $\approx 4\times 4\times4$~mm$^{3}$ single-crystal
sample with a mosaic spread of $\approx25'$. In the first set of
measurements on H4M (experiment I) the $H$-$T$ phase diagram was
determined using neutrons of incident energy $E_{i}=14.7$~meV with a
$(20'-40'-20'-40')$ collimation setup and two PG filters. The second set
of measurements was done on H9 (experiment II) and the field-dependence
of the magnetic propagation vector was studied using
$(60'-40'-60'-\text{sample}-15'-80')$ collimations, an $E_{i}=4.6$~meV
neutron beam and a Be-filter in front of the sample. In both experiments
the use of a pumped $^{4}$He cryomagnet allowed us to work in the
temperature range 1.3--5~K and magnetic fields up to 6.5~T. The sample
was always mounted with the $(h,k,0)$ zone in the scattering
(horizontal) plane and the field was applied along the vertical
direction ($c$-axis of the crystal).

The sample we used in preliminary experiments shattered when it was
cooled down in the cryomagnet for the second time. We therefore tried to
mount the new crystal as strain-free as possible. The new sample was
wrapped
in aluminum foil that was attached to a thin Al-plate. This technique
has one serious drawback. As will be explained below, a good alignment
of the crystallographic $c$ axis with the magnetic field is crucial.
Unfortunately, the described mounting procedure does not allow to
maintain an alignment of better than $\approx 1^{\circ}$. Before the
sample is put into the cryostat, an almost perfect alignment is
achieved. It is upon cooling to base temperature that the undesirable
misalignment occurs. In experiment I the $(1,-1,0)$ and $(1,1,0)$
crystallographic directions formed angles of $1.5^{\circ}$ and
$1^{\circ}$ with the horizontal plane, respectively, as measured {\it in
situ} at low temperature. In experiment II these angles were $1^{\circ}$
and less than $0.3^{\circ}$, respectively.

\section{Experimental results}

\subsection{Magnetization}
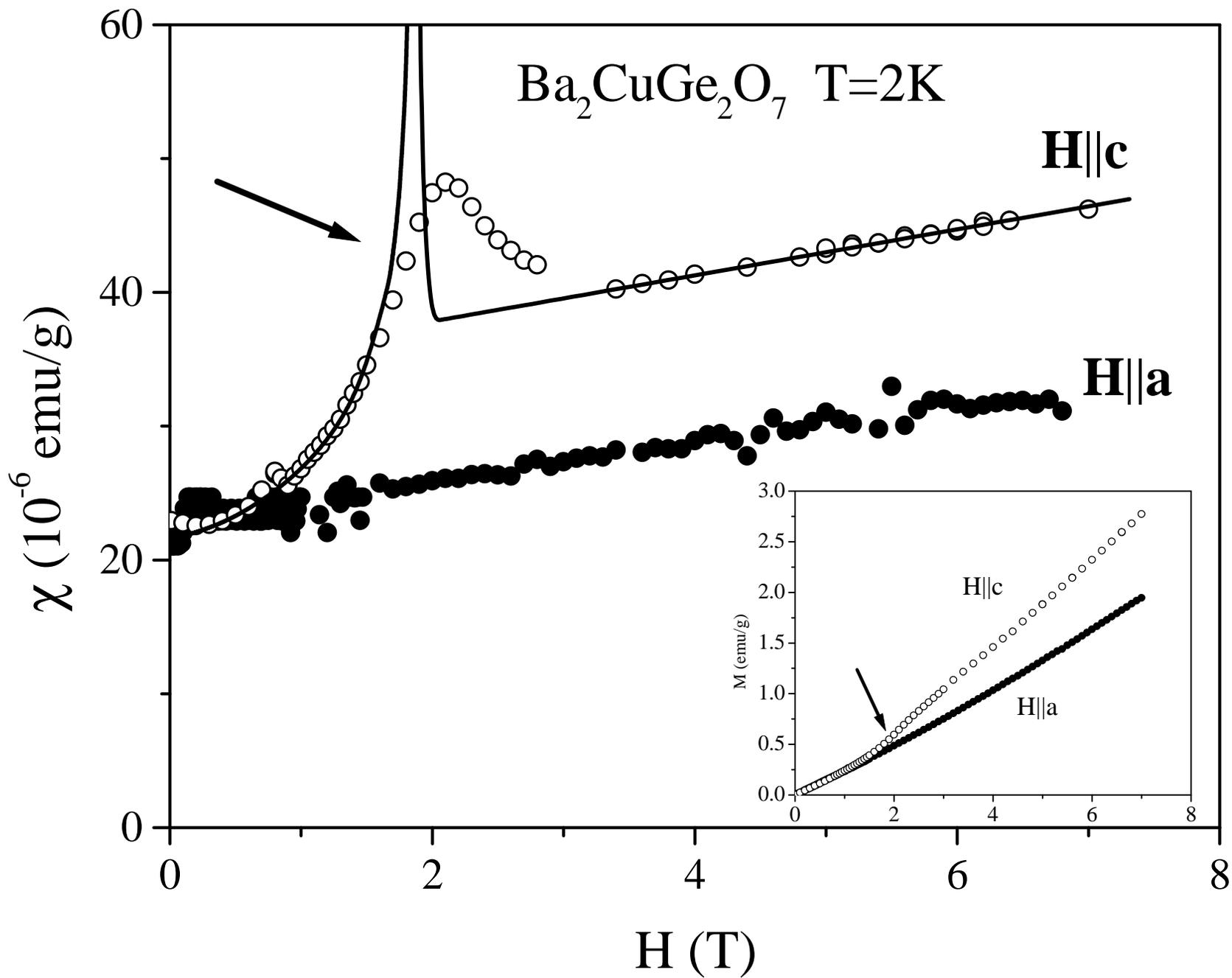
\begin{figure}
\caption{Insert: Magnetization of single crystal Ba$_{2}$CuGe$_{2}$O$_{7}$
measured as a function of magnetic field applied along the $c$ and $a$
crystallographic axes. Main panel: Field-dependence of magnetic
susceptibility $\chi=d M/ dH$ deduced from magnetization curves shown in
the insert. The solid line represents a fit to the data, as described in
the text.}
\label{mag}
\end{figure}

The first evidence of a field-induced magnetic phase transition in
Ba$_2$CuGe$_2$O$_7$ was found in bulk magnetic measurements. The
longitudinal magnetization data collected at $T=2$~K is plotted against
magnetic field ${\bf H}$ applied along the $a$ or $c$ crystallographic
axes in the insert in Fig.~\ref{mag}. For ${\bf H}||{\bf a}$ no
anomalies are observed. In contrast, when the field is applied along
$c$, $M(H)$ has a broad step-like feature around $H=2$~T. The anomaly is
best seen in the plot $\chi(H)$ that was obtained by numerically
differentiating the experimental magnetization curve (Fig.~\ref{mag},
main panel).

\subsection{Neutron diffraction}

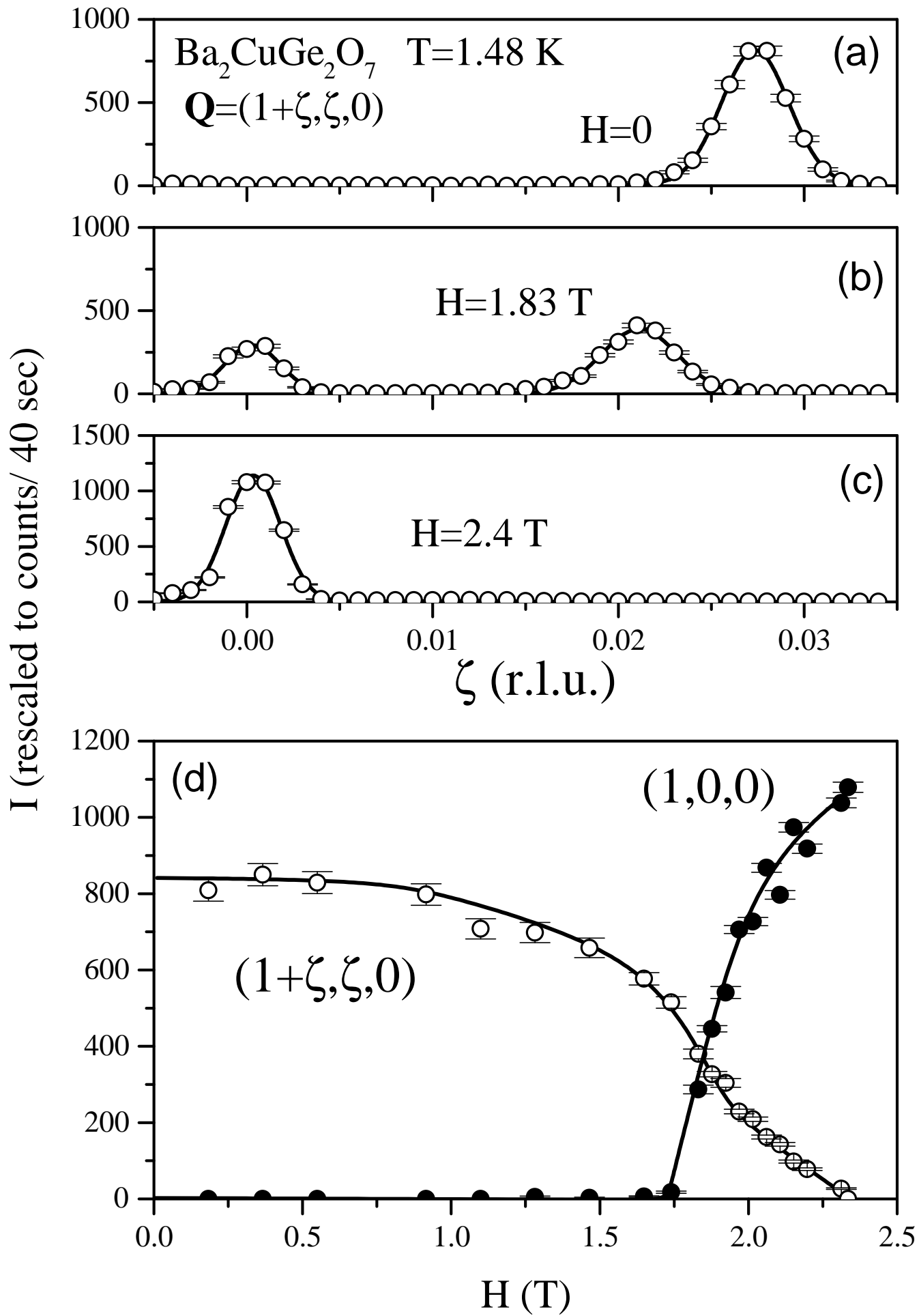
\begin{figure}
\caption{a-c) Elastic scans measured along the $(1+\zeta,\zeta,0)$
direction at $T=1.4$~K for several values of magnetic field applied
along the $c$ axis of the crystal. d) Field-dependence of the intensity
in the $(1,0,0)$ N\'{e}el peak (solid symbols) and the
$(1+\zeta,\zeta,0)$ satellite (open circles).}
\label{exdata1}
\end{figure}

The phase transition is best observed in neutron diffraction
experiments. Figure~\ref{exdata1} (experiment II) shows some elastic
scans along the $(1+\zeta,\zeta,0)$ direction measured in
Ba$_2$CuGe$_2$O$_7$ at $T=1.4$~K. The different scans correspond to
different values of magnetic field applied along the $c$-axis. In zero
field [Fig.~\ref{exdata1}(a)] the magnetic peak is positioned at
$\zeta=0.027$.\cite{Zheludev96-BACUGEO} The incommensurability parameter
$\zeta$ decreases gradually with increasing $H$. Around $H=1.8$~T an
additional peak appears at the N\'{e}el point $(1,0,0)$. At the same
time the intensity of the satellite at $(1+\zeta,\zeta,0)$ starts to
decrease rapidly and the rate at which $\zeta$ changes with field is
increased [Fig.~\ref{exdata1}(b)]. On further increasing the field the
central peak gains intensity and the satellite eventually vanishes at
some critical field $H_{c}\approx 2.3$~T [Fig.~\ref{exdata1}(c)]. At
higher fields only the N\'{e}el (commensurate) peak is seen, all the way
up to the highest field available experimentally. The intensities of the
$(1+\zeta,\zeta,0)$ satellite and the $(1,0,0)$ peak are plotted against
temperature in Fig.~\ref{exdata1}(d). Measured $\zeta (T)$ is shown in
solid circles in Fig.~\ref{good}.

\subsection{Domains and sample alignment}
The field-dependent behavior is {\it extremely} sensitive to the
alignment of the crystallographic $c$ axis with the applied field. The
limitations of the
sample-mounting technique employed, together with technical
impossibility of adjusting the alignment {\it in situ} at low
temperature, are therefore a serious experimental complication. This
problem is closely linked to the issue of magnetic domains. In the
tetragonal symmetry two spiral structures with propagation vectors $(1+
\zeta,\zeta,0)$ and $(1+ \zeta,-\zeta,0)$ are allowed. Zero-field cooling
produces domains of equal volume. If the sample is cooled in an $H=1$~T
field, that is removed only at low temperature, only one domain is
present. Since we are able to exactly measure (but not adjust!) the
orientation of the sample {\it in situ}, we could determine that it is
the domain for which the plane of the spiral forms a larger angle with
the applied field that is stabilized. This behavior can easily be
explained. If the plane in which the spins are confined is not parallel
to the external field, the system can gain some Zeeman energy without
sacrificing much D-M or exchange energy: all spins may tilt slightly in
the direction of the field component normal to the spin plane, producing
a cone structure with practically no change in the angles between
neighboring spins. The domain for which the spin plane forms a larger
angle with the applied field is thus energetically more favorable. Note
that in the commensurate (high-field) structure there are also two
possible domain types, with spins in the $(1,\overline{1},0)$ or
$(1,1,0)$ planes.

The data shown in Fig.~\ref{exdata1} were  collected on a field-cooled
sample and the satellites correspond to the more misaligned
$(1+\zeta,\zeta,0)$ domain (domain A, $\approx 1^{\circ}$ tilt). No
satellites were observed in the $(1+\zeta,-\zeta,0)$ domain (domain B,
$< 0.3^{\circ}$ tilt) in any fields in field-cooling experiments.  When
the measurements at $T=1.4$~K were  done on a zero field cooled sample,
we observed that scattering intensities originating from both domains
remain comparable in fields up to $\approx 1.7$~T. At higher fields the
N\'{e}el peak appears and domain B is rapidly destroyed. Applying a
sufficiently large field thus brings us back to the situation where the
entire crystal is a single magnetic domain.

\subsection{Phase diagram}
The experimental $H-T$ phase diagram for Ba$_2$CuGe$_2$O$_7$ is shown in
Fig.~\ref{phase}. The I-O-II line corresponds to the appearance of the
magnetic reflection at the commensurate $(1,0,0)$ position. The III-O-IV
line shows where the satellite peaks disappear. Both the O-II and the
O-IV lines were measured in a field-cooled sample and the misalignment
of the corresponding domain was $\approx 1.5^{\circ}$ (experiment I).

In the high-field phase the structure is commensurate, with a
propagation vector $(1,0,0)$. This may be the signature of N\'{e}el
order or (see introduction) a canted ``weak ferromagnetic'' structure.
In the low-field region the structure is incommensurate with a
propagation vector $(1+\zeta,\zeta,0)$. Only one domain is present
since, as mentioned, field-cooling was used. Note that except for the
incommensurability of the low-field phase, the phase diagram strongly
resembles that of an easy-axis Heisenberg antiferromagnet in external
field applied along the easy axis.

In the II-O-IV region the N\'{e}el peak coexists with the incommensurate
satellite. It would seem that in the misaligned domain the field-induced
transition is of first order and that the II-O-IV region corresponds to
a mixed-phase state. Indeed, at $T=1.4$~K we have observed some
field-hysteresis: on sweeping the external field down from $H=4$~T the
N\'{e}el component vanishes and the satellite appears at slightly lower
fields, than in the case when the $H$ is gradually increased from zero.
We believe that in a perfectly aligned sample the transition must be of
2nd order, in which case the II-O-IV region is replaced by a single
line.

\subsection{Measurements of $\zeta(H)$}
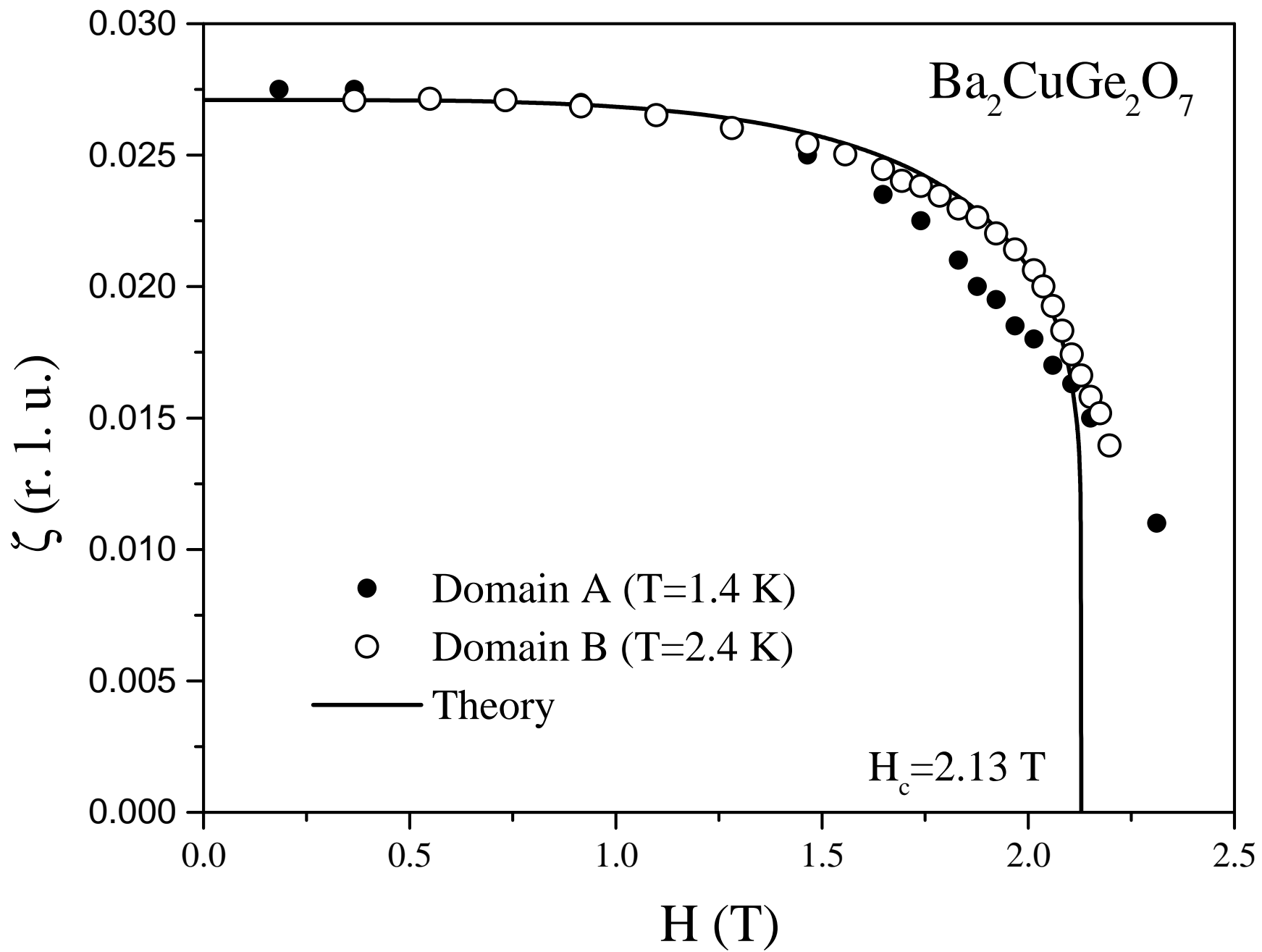
\begin{figure}
\caption{Measured field dependence of the magnetic propagation vector
$\zeta$ in the incommensurate phase of Ba$_{2}$CuGe$_{2}$O$_{7}$. The data
shown by solid circles correspond to a domain misaligned by $\approx
1^{\circ}$ with respect to the applied field, and were collected on a
field-cooled
sample at $T=1.4$~K. Open circles are data points measured at $T=2.4$~K in
an almost perfectly aligned domain, obtained by zero-field-cooling the
sample from $T=5$~K. The solid line is a fit with a theoretical curve
described in the Theory section.}
\label{good}
\end{figure}

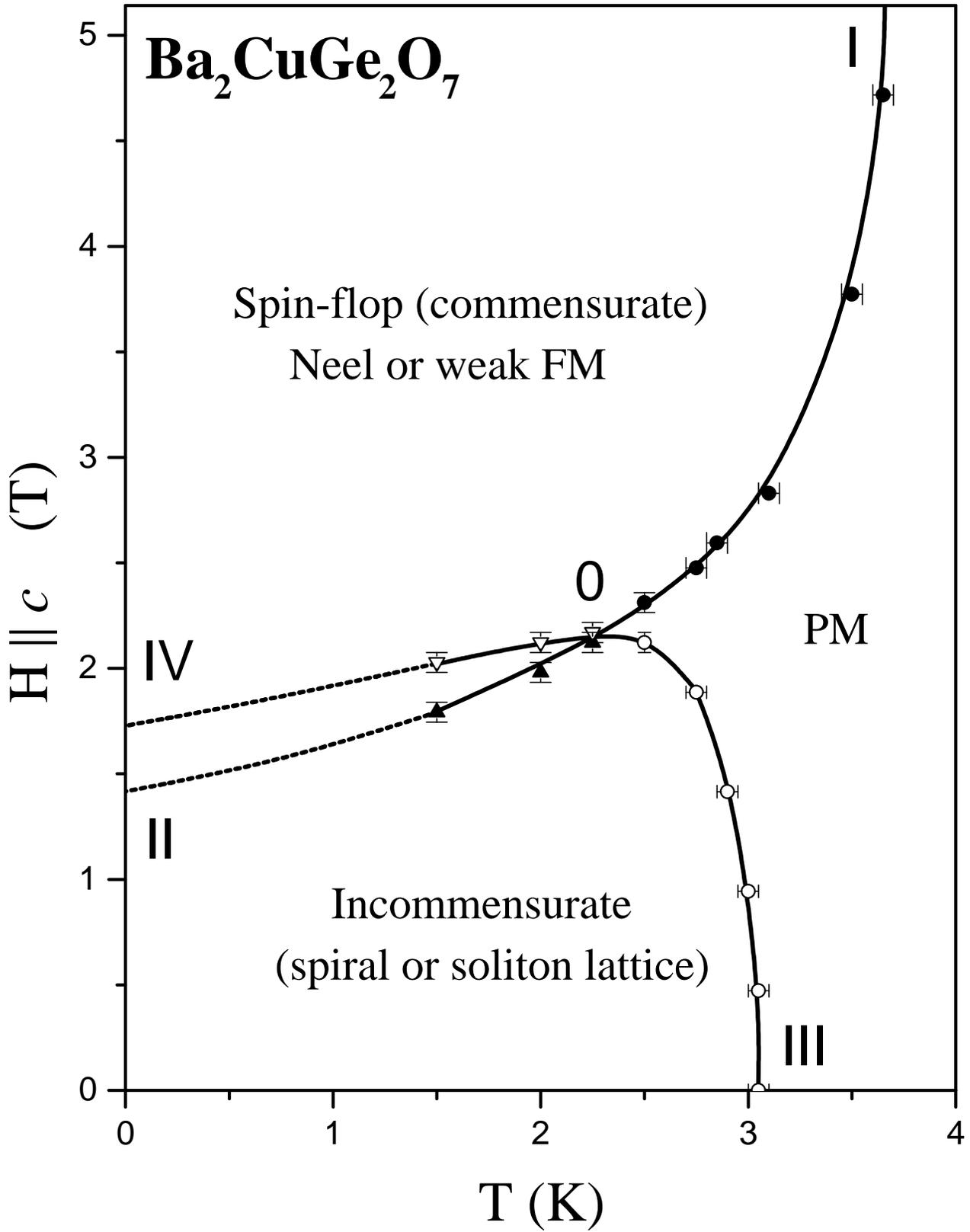
\begin{figure}
\caption{Measured~magnetic~phase~diagram~of Ba$_{2}$CuGe$_{2}$O$_{7}$. The
field is applied along the $c$-axis of the crystal.}
\label{phase}
\end{figure}

Accurate measurements of the field-dependence of $\zeta$ were performed
in experiment II where one domain was tilted by $\approx 1^{\circ}$
(domain A) and the other almost perfectly aligned (domain B).
Measurements for domain A could easily be done at base temperature
(1.4~K) in a field-cooled sample (Fig.~\ref{good}, solid circles).
Additional data for domain A were collected at $T=2.4$~K. At this
temperature $\zeta(H)$ was found to follow the same curve as at
$T=1.4$~K to within experimental error.

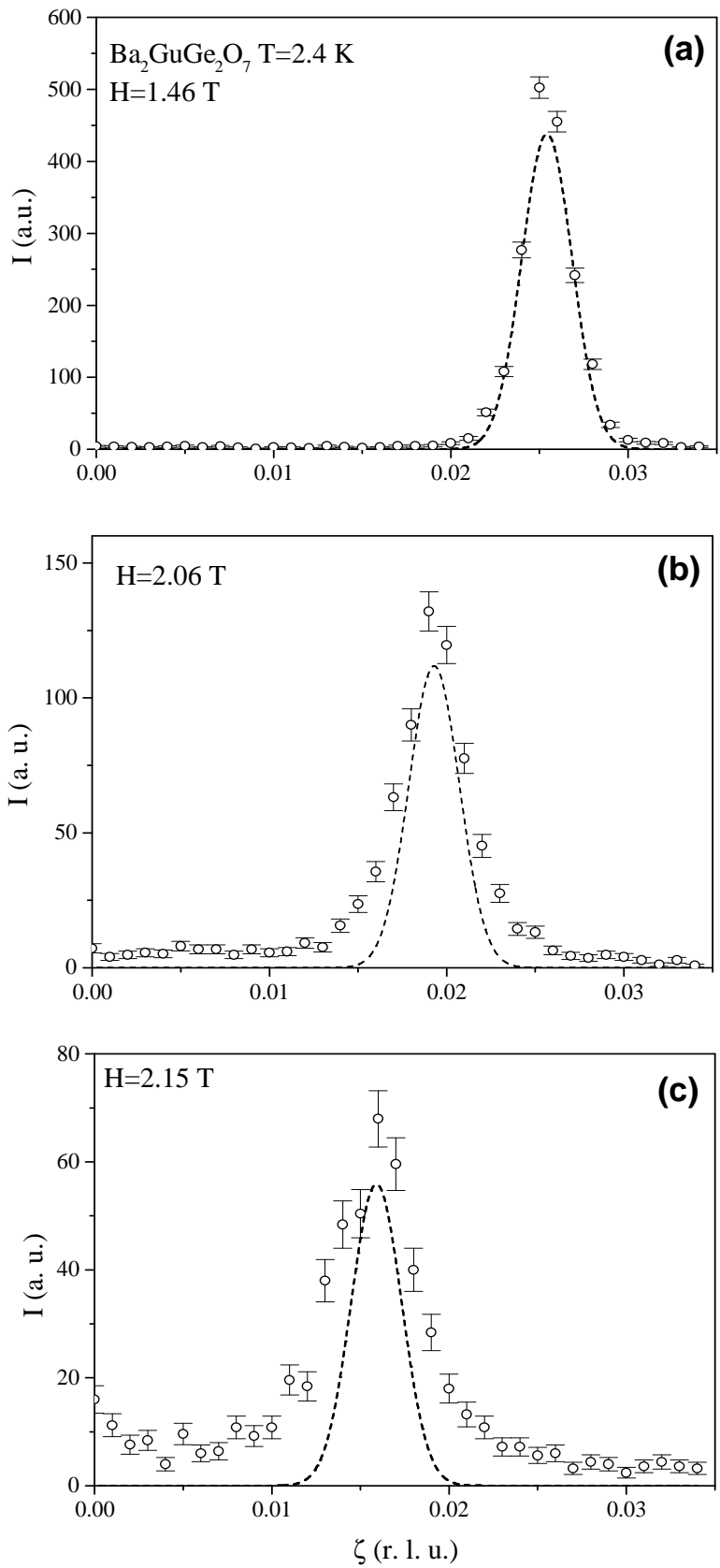
\begin{figure}
\caption{Elastic scans measured along the $(1+\zeta,-\zeta,0)$ direction in
Ba$_{2}$CuGe$_{2}$O$_{7}$
at $T=2.4$~K for several values of magnetic field applied along the $c$
axis of the crystal. For this domain the $(1,0,0)$ component does not
appear
before the
critical field is reached.}
\label{exdata2}
\end{figure}

For domain B experiments in the interesting field region, where $\zeta$
starts to decrease rapidly with increasing $H$, could not be performed
at low temperature: the domain itself is destroyed in fields higher than
$\approx 1.7$~T (see above). Nevertheless, $\zeta(H)$ for domain B could
be measured in a zero-field-cooled sample at $T=2.4$~K, i.e., just at
the temperature of the critical point O in the phase diagram. At this
temperature the B-domain survives up to the critical field and the
N\'{e}el component does not appear before the $(1+\zeta, -\zeta, 0)$
peak vanishes. Apparently even at $2.4$~K domain wall pinning is
sufficiently strong and kinetics sufficiently slow to make the
energetically less favorable (better aligned) domain metastable even in
high fields. This does not happen at low temperature: the commensurate
structure emerges in a 1st-order transition and presumably the
thermodynamic force that destroys the better aligned domain is larger.
Some typical scans for domain B at $T=2.4$~K are shown in
Fig.~\ref{exdata2}. The dashed line shows experimental $Q$-resolution.
The $(1+\zeta, -\zeta, 0)$ satellite is resolution-limited at $H<1.95$~T
[Fig.~\ref{exdata2}(a)] and starts to gradually broaden at higher fields
[Fig.~\ref{exdata2}(b,c)]. The experimental $\zeta(H)$ is plotted in
open circles in Fig.~\ref{good}. We clearly see the sensitivity of the
system to even a minor tilt of the applied field relative to the $c$
axis. While at low fields $\zeta(H)$ for domains A and B coincide, there
is a substantial discrepancy near the transition point.

\subsection{Critical indexes}
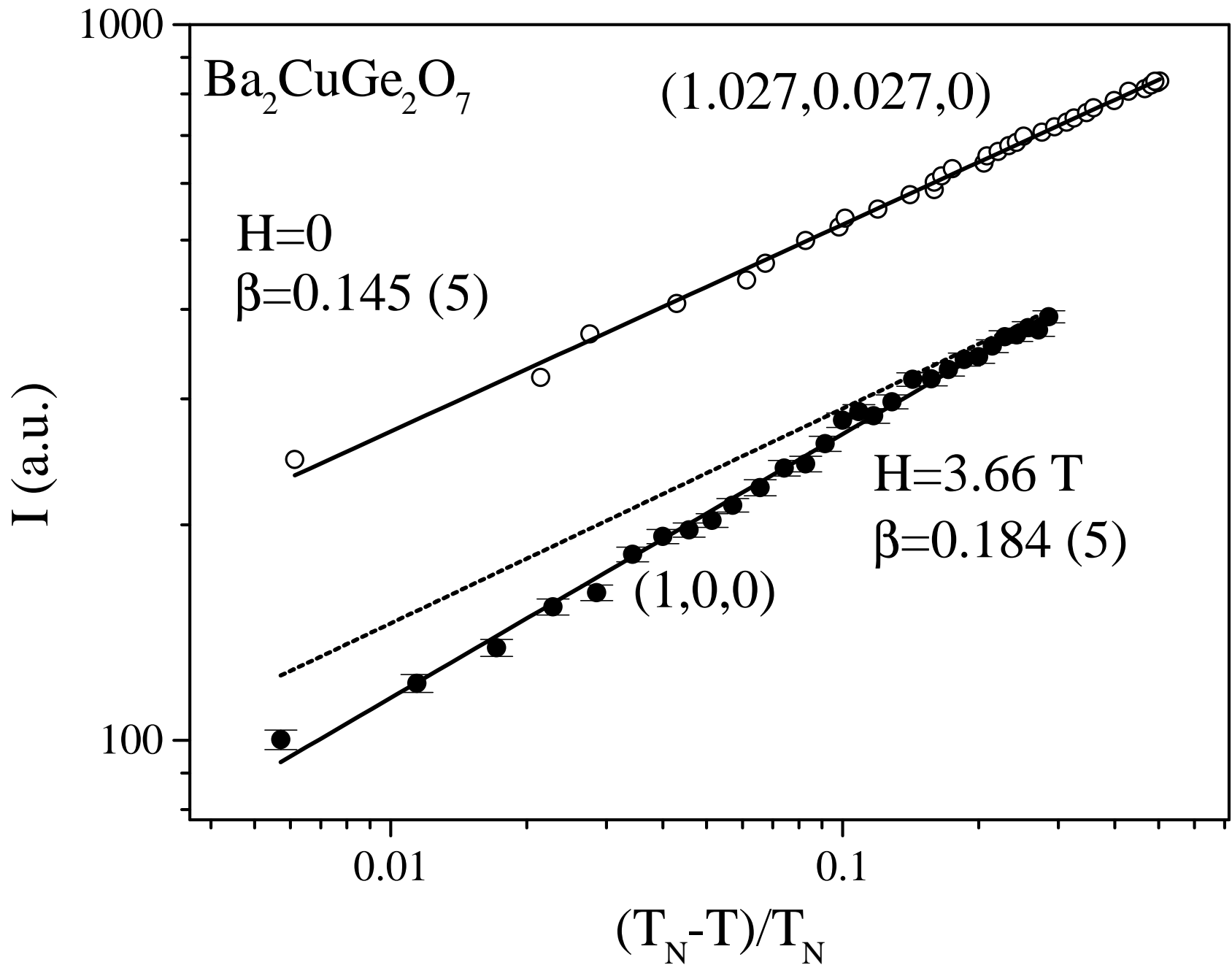
\begin{figure}
\caption{Measured Bragg intensity of the $(1,0,0)$ (open circles,
$H=3.66$T, $T_{N}=3.5$~K) and
$(1+\zeta,\zeta,0)$ (solid circles, $H=0$,$T_{N}=3.2$~K) magnetic Bragg
reflections. The solid lines lines are power-law fits to the data. The
dashed line is an aid to the discussion in the text.}
\label{critical}
\end{figure}

In our previous work we have reported measurements of the
order-parameter critical exponent $\beta=0.145(0.005)$ for the
zero-field transition from paramagnetic (PM) to spiral states. Having
discovered the commensurate phase, we have in addition determined
$\beta$ for the PM $\rightarrow$ commensurate transition at $H=3.66T$.
The intensity of the $(1,0,0)$ magnetic reflection is plotted against
$(T_{N}-T)/T_{N}$ in a log-log plot in Fig.~\ref{critical} (open
circles). Fitting the data with a power-law dependence we obtain
$\beta_{H=4T}=0.185(0.005)$. The previously measured temperature
dependence of the $(1.0273,0.0273,0)$ Bragg intensity at $H=0$ is shown
in solid symbols in Fig.~\ref{critical}. Although the experimentally
determined critical indexes are rather close to each other, in our data
we clearly see that $\beta=0.145$ is incompatible with the 4~T
measurement (Fig.~\ref{critical}, dashed line).

\section{Theory}
In the introduction section we have qualitatively shown how D-M
interactions stabilize the magnetic spiral in Ba$_2$CuGe$_2$O$_7$ and
why applying a field along the $c$ axis of the crystal should result in
a commensurate structure. We shall now develop a quantitative
description of a D-M spin-spiral in an external magnetic field. The
general expression for the free energy of such system was obtained by
Dzyaloshinskii,\cite{Dzya65} who also predicted the CI transition and
obtained analytical expressions for the critical properties [at
$(H_c-H)/H << 1$]. In the present work we extend Dzyaloshinskii's
approach and derive analytical expressions for the period and the
uniform magnetization of the spiral valid throughout the phase diagram.
We start by considering a simple, yet illustrative, case of a classical
spin chain at $T=0$.

\subsection{Classical spin chain with D-M interactions.}
Let us consider the 1-D case: a uniform chain of classical spins with
isotropic (Heisenberg) antiferromagnetic exchange. In addition, we
include a Dzyaloshinskii energy term with the ${\bf D}$ vector pointing
along the arbitrary chosen $y$-axis. The system is then characterized by
the following Hamiltonian (energy functional):
\begin{equation}
{\cal H}=2J \sum_{n}{\bf S}_{n}{\bf S}_{n+1}+ D \sum_{n}({\bf
S}_{n}\times {\bf
S}_{n+1})_{y}
\end{equation}
It is straightforward to show that to take full advantage of the
Dzyaloshinskii term, the spins have to be confined to the $(x,z)$ plane.
The total interaction energy of a pair of spins is given by ${\cal E}=2J
S^2 \cos \phi + DS^2 \sin \phi=S^2\sqrt{ 4 J^2 +D^2} \cos(\phi
-\alpha)$, where $\phi$ is the angle between the spins, and $\alpha =
\arctan D/2J\ll 1$, provided $J \gg D$. The classical ground state is
therefore an AF spiral, where the angle between subsequent spins is
equal to $\pi+\alpha$.

Physically interesting behavior occurs when an external magnetic field
is applied perpendicular to the Dzyaloshinskii vector ${\bf D}$, e.g.,
along $z$-axis. In this case one has to take into account the Zeeman
energy $-H g\mu_{B} \sum_{n}S_{z}$, where $g$ is the gyromagnetic ratio
and $\mu_{B}$ is the Bohr magneton. Denoting by $\phi_n$ the angle the
$n$-th spin ${\bf S}_n$ forms with the $z$-axis, we can rewrite the
energy functional as:
\begin{equation}
{\cal H}=\sum_n [\tilde{J}\cos(\phi_{n+1}-\phi_{n}-\alpha) -\tilde{H} \cos
\phi_n],
\label{Ham}
\end{equation}
Where $\tilde{J}=S^2 \sqrt{ 4J^2+D^2}$ and $\tilde{H}=H g\mu_B S$.

In Eq.(\ref{Ham}) we clearly see the competing terms that drive the CI
transition. The $J$-term favors an incommensurate (spiral) structure,
where the phases $\phi_{n}$ of subsequent spins differ by $\pi+\alpha$.
The effect of the magnetic field is more subtle. For
$\tilde{H}\gg\tilde{J}$ the spins tend to align parallel to the field.
For $\tilde{H}\ll\tilde{J}$ however, the system still can gain some Zeeman
energy without sacrificing much of its exchange energy. This is achieved
in a spin-flop state, where all spins are roughly perpendicular to the
magnetic field and slightly tilted in its direction. The external field
thus favors a commensurate spin-flop structure, but it can only be
realized by sacrificing some Dzyaloshinskii energy. Indeed, the energy
difference between the spin-flop configuration and the spiral state is
$\tilde{J} (1-\cos \alpha)-\tilde{H}^2/8\tilde{J}$. The spin flop state
becomes energetically favorable in fields $\tilde{H}>\tilde{H}_c \sim
2\alpha \tilde{J}$. This is the simplified physical picture already
discussed in the introduction section. The above crude arguments does
not take into account the {\it distortion} of the spiral by the applied
field. The exact result for our model, $\tilde{H}_c=\pi \tilde{J}
\alpha$ is derived below.

\subsection{Exact ground state in the continuous limit}
It is convenient to change angle variables so that
\begin{equation}
\phi_n=\pi n + \theta _n
\end{equation}
The energy functional in this notation becomes
\begin{equation}
{\cal H}= \sum_n\left[-\tilde{J}\cos(\theta_{n+1}-\theta_{n}-\alpha) -
(-1)^n \tilde{H}\cos \theta_n\right].\label{az1}
\end{equation}
The advantage of the new set of variables is that assuming $\alpha = 2 \pi
\zeta << 1$ and $\tilde{H} << \tilde{J}$, the phase difference
$\theta_{n+1}-\theta_{n}-\alpha << 1$, so we can safely replace the 1st
term in Eq.(\ref{az1}) with $-\tilde{J} + \tilde{J}
(\theta_{n+1}-\theta_{n}-\alpha)^2/2$.

The ground state satisfies the extremal conditions $\partial{\cal
H}/\partial{\theta_{n}}=0$, which gives us the following set of
equations:
\begin{equation}
\theta_{n+1}+\theta_{n-1}-2\theta_{n}={\tilde{H} \over \tilde{J}} (-1)^n
\sin \theta_n
\label{e.der}
\end{equation}

These equations can be solved in the continuous limit. We shall look for a
solution in the form $\theta_{n}=\Theta(n)+(-1)^n \vartheta (n)$, where
both $\Theta(n)$ and $\vartheta(n)$ are functions that only slowly change
on the scale of a single lattice spacing. From Eq. (\ref{e.der}) we obtain
the extremal condition that should be satisfied in an energy minimum:
\begin{eqnarray}
 {d^2 \Theta \over dn^2}=-{\tilde{H}^2 \over 8\tilde{J}^2}
\sin 2\Theta  \label{SG}\\
 \vartheta  =-{\tilde{H} \over 4\tilde{J}} \sin \Theta \label{SGG}
\end{eqnarray}

The first of these equations is the famous sine-Gordon equation, which
allows for exact soliton solutions. The trivial ``no-soliton'' solution
to Eqs.~(\ref{SG},\ref{SGG})  is $\Theta (n)=\pi /2$, and $\vartheta
(n)=
-\tilde{H}/4\tilde{J}$. This is precisely the spin-flop phase,
realized in strong enough magnetic field.
Less trivial is the one-soliton solution:
\begin{eqnarray}
\Theta(n)=2\arctan [ \exp (n\tilde{H}/2\tilde{J})]-\pi/2 \label{e.sg}\\
\vartheta (n)=-\tilde{H}/4\tilde{J} \tanh (n\tilde{H}/2\tilde{J}) \nonumber
\end{eqnarray}
which has boundary conditions
$\Theta(-\infty)=-\pi/2$ and $\Theta(+\infty)=\pi/2$.

For $\alpha=0$ the energy of the one-soliton state will be always higher
than that of the soliton vacuum (spin-flop state). In the general case,
substituting (\ref{e.sg}) into the energy functional (\ref{az1}), we
find the energy difference between the one-soliton and soliton-free
state: ${\cal E}_{1}-{\cal E}_{0}=\tilde{H}-\tilde{J}\pi \alpha$. We see
that for $\tilde{H}<\tilde{H}_{c}$, where
\begin{eqnarray}
\tilde{H}_c=\pi \alpha \tilde{J}=2 \pi ^2 \zeta \tilde{J} \qquad
; \nonumber \\
H_c=4 \pi ^2 \zeta { J S \over g \mu _B} \qquad , \label{Hc}
\end{eqnarray}
the energy of a single soliton is negative and solitons spontaneously
``condense''. This process will eventually saturate, due to the mutual
repulsion of solitons, and the ground state will be a periodic ``soliton
lattice''.

At the non-negligible soliton density the interaction starts to change
the shape of individual solitons and has to be taken into account
explicitly. Fortunately the general solution of the sine-Gordon equation
is known:
\begin{eqnarray}
\int_0^{\Theta (n)} {\beta dx \over \sqrt{1-\beta^2 \sin ^2 x}}
={n\tilde{H} \over 2\tilde{J}} \qquad ; \\
\Theta(n)={\rm am}( n {\tilde{H} \over 2\tilde{J} \beta}, \beta) \qquad
\label{s1};
\\
\vartheta (n)=-{\tilde{H} \over 4\tilde{J}} \sin \Theta(n) = -{\tilde{H}
\over 4\tilde{J}}
{\rm sn}( n{\tilde{H} \over 2\tilde{J} \beta}, \beta) \qquad \label{s2},
\end{eqnarray}
where $am(x, \beta)$ and $sn(x, \beta)$ are Jacobi elliptic functions of
modulus $\beta$. Substituting this solution into the energy functional
(\ref{az1}) we obtain the interaction energy per spin:
\begin{equation}
{\cal E}={(g \mu_B H)^2 \over 16 J}
\left[-{1 \over \beta^2} + {2 E(\beta) \over \beta^2 K(\beta)}-
{2 H_c \over \beta H K(\beta)}\right] \qquad ,
\label{e.en_beta}
\end{equation}
where K and E are the complete elliptic integrals of the first and
second kind, respectively.

Equation (\ref{e.en_beta}), as well as the expression for $H_{c}$
[Eq.~(\ref{Hc})] were derived in Ref.~\onlinecite{Dzya65}. In the
present work we go one step further and determine which of the solutions
given by Eqs.~(\ref{s1},\ref{s2}) indeed correspond to the ground state
at any particular $H<H_{c}$. Unlike that in Ref.~\onlinecite{Dzya65},
our treatment is not limited to the narrow critical region $(H_c-H)/H_c
<<1$. Taking the partial derivative of the energy functional
(\ref{e.en_beta}) with respect to $\beta$, and substituting identities
for derivatives of elliptic integrals \cite{Gradstein80} after some
algebra we get:
\begin{eqnarray}
{H \over H_c} = {\beta \over E( \beta)} \qquad .
\label{e.x} \\
{\cal E}=-{(g \mu_B H)^2 \over 16 J} {1 \over \beta ^2}
\label{e.en_min}
\end{eqnarray}
Equation (\ref{e.x}) is to be solved with respect to $\beta$ for any
given $H/H_c$. $\zeta(H)$ can then be expressed in terms of $\beta$:
\begin{equation}
{\zeta(H) \over \zeta(0)}={H \over H_c} {\pi ^2
\over 4
\beta  K( \beta)}={\pi^2 \over 4 E(\beta) K(\beta)} \qquad .
\label{e.zeta}
\end{equation}
Equations (\ref{e.x}) and (\ref{e.zeta}) provide us with $\zeta(H)$ in a
parametric form. The ratio $\zeta(H)/\zeta(0)$ is plotted against
$H/H_{c}$ in Fig.~\ref{theory}. The insert schematically shows the spin
structure for $H=0$, $0<H<H_{c}$, and $H>H_{c}$.

At low density soliton repulsion is exponentially small and the system
close to the transition point is extremely sensitive to any
perturbations of the original Hamiltonian, that will tend to pin down
the soliton lattice in one of the metastable configurations. As
$H\rightarrow H_{c}-0$, the soliton density decreases as $1/|\ln
(H_c-H)|$, i.e very rapidly.\cite{Dzya65} The transition is thus almost
first order [the critical exponent $\delta=0$ in $\zeta(H) \sim (H_c
-H)^{\delta}$].

The soliton lattice at $H\ne 0$, unlike the pure sinusoidal spiral
at $H=0$, has higher Fourier harmonics at $\zeta_{2n+1}=\pm (2n+1)
\zeta (H)$. The knowledge of the exact ground state enables us to
calculate these harmonics analytically. For the Bragg intensity (square
of the amplitude) of the strongest 3rd harmonic, for example, one gets
\begin{eqnarray}
{I_3(H) \over I_1(H)} &=& {q^2(1-q^2+q^4) \over (1+q^2+q^4)^2} \qquad ;\\
{I_3(H) \over I_1(0)} &=& {2 \pi ^2 ( q^3/(1-q^3)^2+q^3/(1+q^3)^2)
\over
\beta^2 K(\beta)^2}\\ \qquad,{\rm where} \quad
q &=& \exp ( -\pi {K(\sqrt{1-\beta^2}) \over K(\beta)}) 
\qquad . \nonumber
\end{eqnarray}
This ratio of intensities is plotted against $\zeta(H)/\zeta(0)$ in
Fig.~\ref{harmonic}. Except very close to $H_c$ all harmonics are
much weaker than the first one.

As we derived a parametric expression for $\zeta(H)$, so we can also obtain
an exact parametric form for the uniform magnetization $M=-\partial
{\cal E}/ \partial H$. From Eqs. (\ref{e.x},\ref{e.en_min}), and the
identities for derivatives of elliptic integrals \cite{Gradstein80} one
gets:
\begin{equation}
M= {(g \mu _B)^2 H \over 8 J}{1 \over \beta ^2}
\left(1-
{E( \beta) \over K(\beta)}\right) \qquad . \label{e.mag}
\end{equation}
The magnetization curve is continuous at the critical field. On the
other hand, its derivative with respect to $H$ (the field dependent
susceptibility) diverges when $H_c$ is approached from below.

\subsection{Generalization to finite temperatures and higher dimensions}
The results obtained in the previous chapter can be easily generalized
to the cases of higher system dimensionality and finite temperatures. We
follow the approach developed by Dzyaloshinskii in Ref.
\onlinecite{Dzya65} and introduce the expression for the free energy of
a general almost antiferromagnetic spiral in an external
magnetic field. We generalize the results derived for the 1-D chain
in the previous section to obtain analytical expressions for the
period and the
magnetization of the structure valid not only in the critical region,
but troughout the phase diagram.

Since the period of the spiral is long, nearest neighbor magnetic
moments are almost opposite to each other. This enables us to
approximately describe the structure in terms of position dependent
staggered magnetization ${\bf L}({\bf r})$. To take full advantage of
the D-M interaction, ${\bf L}({\bf r})$ must be perpendicular to the D-M
vector, and therefore be confined to the $x-z$ plane. We further assume
that the magnitude $|{\bf L}({\bf r})|$ does not depend on ${\bf r}$.
This simplification is justified since the Zeeman energy, which at
$H_{c}$ is of order of D-M interaction energy, is much smaller than the
energy of isotropic exchange. The local magnetic structure is thus fully
defined by the angle $\theta ({\bf r})$, that ${\bf L}({\bf r})$ forms
with the $z$-axis. The total ``elastic'' free energy per single Cu
plane, associated with spin-spin interactions, can then be written as:
\begin{eqnarray}
F_{elast}= \frac{1}{2}\rho _s (T) \int dxdy
\left( \left({\partial \theta ({\bf r}) \over
\partial x}  - {\alpha \over
\Lambda} \right)^2+ \right. \nonumber \\
\left. + \left({\partial \theta ({\bf r}) \over
\partial y} \right)^2 +  
\gamma \left( {\partial \theta ({\bf r}) \over
\partial z} \right)^2 \right) \label{ela}.
\end{eqnarray}
Equation (\ref{ela}) is given in the form suitable for describing a
single Cu plane in Ba$_{2}$CuGe$_{2}$O$_{7}$. $\rho _s(T)$ is the
so-called spin stiffness at given temperature, $\Lambda=\Lambda_{ab}$ is
the in-plane nearest neighbor Cu-Cu distance, and $\gamma$ is the spin
stiffness anisotropy factor, defined as $\gamma (\Lambda_c/
\Lambda_{ab})^2 = |J_c|/|J_{ab}| \approx 1/37$ in
Ba$_{2}$CuGe$_{2}$O$_{7}$. Coordinates $x$, $y$, and $z$ run along
(1,1,0), (1,$\overline{1}$,0), and (0,0,1) directions, respectively. The
spin stiffness at zero temperature is given by $\rho_s (0)= 2J
S^2$.\cite{Chakravarty89} As $T$ approaches $T_N$ the spin stiffness
decreases as $|{\bf L}({\bf r})|^2$ and vanishes precisely at $T_N$.

In taking into account the external magnetic field, it is convenient to
introduce the quantities $\chi_{\|}(T)$ and $\chi_{\perp}(T)$, the
magnetic susceptibilities of the system with respect to a field that
rotates along with the spiral structure and is always parallel or
perpendicular to the local staggered magnetization, respectively. As in
a usual antiferromagnet, these susceptibilities do not diverge at $T_N$.
In the paramagnetic phase (at $T>T_N$), $\chi_{\|}(T)=\chi_{\perp}(T)$.
At zero temperature the classical result is $\chi_{\perp} (0)= (g
\mu_B)^2/(8d J \Lambda ^2 )$,\cite{Chakravarty89} and $\chi_{\|} (0)=0$.

The full expression for the free energy density in the presence of
magnetic field $H$ applied along $z$-direction is:
\begin{eqnarray}
F= F_{elastic} -
{\chi_{\perp}H^2 \sin ^2 \theta ({\bf r})+
\chi_{\|}H^2 \cos ^2 \theta ({\bf r}) \over 2}  =
\nonumber \\
={\rho _s \over 2}
\left( \left({\partial \theta ({\bf r}) \over
\partial x}  - {\alpha \over
\Lambda}\right)^2+\left({\partial \theta ({\bf r}) \over
\partial y}\right)^2 +  \right. \nonumber \\
\left.
\gamma \left( {\partial \theta ({\bf r}) \over
\partial z} \right)^2 \right)
- {(\chi_{\perp}-\chi_{\|}) H^2 \over 2}
 \sin ^2 \theta ({\bf r}) - {\chi_{\|} H^2 \over 2}.
\label{e.energy}
\end{eqnarray}
At $T=0$ and $d=1$ Eq.(\ref{e.energy}) coincides with the expression for
the energy of a spin chain derived in the previous section.

The equilibrium configuration of $\theta ({\bf r})$ minimizes the free
energy and therefore satisfies a generalized form of Eq. (\ref{SG}):
\begin{equation}
 {\partial ^2 \theta \over
 \partial x^2}=-{(\chi_{\perp}-\chi_{\|}) H^2 \over \rho _s}
 \sin \theta \cos \theta = -{1 \over 2 \Gamma^2} \sin 2\theta,
 \label{e.s-g}
\end{equation}
where $\Gamma=(\rho _s/H^2 (\chi_{\perp}-\chi_{\|}))^{1/2}$. The
critical field is now given by
\begin{equation}
 H_{c}=\pi ^2 \zeta(0) \sqrt{\rho_s
 \over (\chi_{\bot}-\chi_{\|}) \Lambda^2 }
 \label{2dhc}
\end{equation}
For a 4-nn AF like Ba$_{2}$CuGe$_{2}$O$_{7}$ at $T=0$ we can use
$\chi_{\perp}=(g \mu_B)^2/16 J \Lambda^2$, $\chi_{\|}=0$, and
$\rho_s=J_{ab} S^2$. Substituting these expressions into Eq.
(\ref{2dhc}) we obtain
\begin{equation}
 H_c=4\sqrt{2}\zeta \pi^2 \frac{JS}{g\mu_{ B}} ,
 \label{estimate}
\end{equation}
a factor of $\sqrt{2}$, compared to the 1-D case [Eq. (\ref{Hc})]. This
factor is due to the fact that in 2 dimensions each spin
is antiferromagnetically
coupled to four, rather that two nearest neighbors, and the overall
structure is stiffer. Unlike antiferromagnetic in-plane interactions,
{\it ferromagnetic} coupling between adjacent Cu-planes in
Ba$_2$CuGe$_2$O$_7$ does not change $\chi_{\perp}$ or the expression for
the critical field.

The temperature dependence of $H_c$ can be easily understood within the
mean-field (MF) approximation. In this framework $\chi_{\bot}$ is
$T$-independent, $\chi_{\|}$ decreases with $T_{N}-T$ and
$\chi_{\bot}=\chi_{||}$ at $T=T_{N}$. The effective strength of exchange
coupling, represented by $JS^2$ in Eq.\ (\ref{2dhc}) goes as the square
of the order parameter, i.e., as $T_{N}-T$ (the MF order-parameter
critical exponent $\beta=0.5$). Substituting these values in
Eq.(\ref{2dhc}), we find that within the MF approximation $H_{c}$ is
independent of temperature. This result is the same as for the spin-flop
field in a conventional easy-axis Heisenberg antiferromagnet.

Finally, we can  generalize the expression (\ref{e.mag}) for the
magnetization.
With new parameters in the sine-Gordon equation we get:
\begin{equation}
        M= \chi_{\|} H + (\chi_{\perp}-\chi_{\|}){ H \over \beta ^2}
        \label{chi}
        \left(1-{E( \beta) \over K(\beta)}\right) \qquad ,
\end{equation}
for $H \leq H_c$.
For $H>H_c$ one has $M=\chi_{\perp} H$.
The field dependent susceptibility $\chi (H)= dM/dH$ is given by:
\begin{equation}
\chi (H)=\chi_{\|}+{(\chi_{\perp}-\chi_{\|}) \over \beta ^2}
\left(1+{E(\beta)^3 \over (1- \beta ^2) K(\beta)^3}-{2 E(\beta)
\over K(\beta)} \right).
\end{equation}
At $H=0$ this expression gives $\chi (0)=(\chi_{\perp}+\chi_{\|})/2$:
the structure is a uniform spiral, which effectively
averages out the susceptibility for all directions in the spin plane.
The susceptibility diverges at $H_c$ as $1/(H_c-H)\ln ^2
(H_c-H)$.\cite{Dzya65} Above $H_c$ it has a constant value equal to
$\chi_{\perp}$.

\section{Discussion}

\subsection{D-M interactions in Ba$_{2}$CuGe$_{2}$O$_{7}$}
We have shown that the spiral spin arrangement in
Ba$_{2}$CuGe$_{2}$O$_{7}$ is due to the in-plane component $D_{y}$ of
the Dzyaloshinskii vector ${\bf D}$ for nearest-neighbor Cu$^{2+}$ ions.
It is rather difficult to experimentally determine whether or not the
out-of-plane component $D_{z}$ is also active. In the spin-flop phase
the predicted canted weak-ferromagnetic structure differs from a
N\'{e}el state only in that it gives rise to additional magnetic peaks
coincident with nuclear Bragg reflections. The latter are much stronger
than any magnetic scattering intensities, and make measurements of the
ferromagnetic component all but impossible. In the spiral phase however,
if $D_{z}\ne 0$, satellites of type $(h+\zeta,k+\zeta,l)$ should be
observable around {\it ferromagnetic} zone-centers, in addition to the
principal magnetic peaks positioned around the AF zone-centers. In
preliminary experiments at $T=1.5$~K we have indeed observed extremely
weak elastic features at reciprocal-space positions $(2+\zeta,\zeta,0)$
and $(1+\zeta,1+\zeta,0)$. So far we have not investigated the
possibility of these features being artifacts due to nuclear-magnetic
double scattering.\cite{moon64} We plan to resolve this uncertainity in
future experiments.

\subsection{The commensurate-incommensurate transition}

Studies of CI phase transitions have a long history, dating back to
pioneering works of Frenkel and Kontorova \cite{Frenkel38} and Frank and
van der Merwe.\cite{Frank49}. Since then CI transitions were discovered
and studied experimentally in a number of such seemingly unrelated
systems as noble gas monolayers adsorbed on graphite
surface\cite{Clarke80}, charge density wave materials\cite{Wilson75},
ferroelectrics\cite{Moudden82} and rare-earth magnets\cite{Koehler72}
(For comprehensive reviews see for example Ref.~\cite{Bak82}). As a
rule, CI transitions result from a competition between two distinct
terms in the Hamiltonian that have different ``built-in'' spatial
periodicities and are often referred to as potential and elastic energy,
respectively. The potential energy by definition favors a structure
commensurate with the crystal lattice. Such is the interaction between
gas atoms and the graphite matrix in intercalated and adsorbed systems.
The elastic term is intrinsic to the system where the transition occurs,
and has a different ``natural'' built-in period. For adatoms on graphite
this term represents their mutual interaction. In our case of a
Heisenberg AF with D-M interactions it is the $\tilde{H}\cos \phi _n$
term in Eq. (\ref{Ham}) that plays the role of an effective potential,
forcing the spins in the plane, and thus favoring a commensurate
structure. The competing elastic term is
$\tilde{J}\cos(\phi_{n+1}-\phi_{n}-\alpha)$, and the ``natural''
periodicity is set by the angle $\alpha$.

In many known realizations of CI transitions, such as adsorbed gas
monolayers, it is the period set by the elastic term that can be varied
in an experiment to drive the transition, whereas both the strength and
the period of the potential remain constant. In other systems, among
them rare-earth magnets, both the elastic term (exchange coupling
between spins) and the potential (magnetic anisotropy) can be changed,
but only indirectly, by varying the temperature. In both cases one
typically observes a ``devil's staircase'' phase diagram (for a review
see for example Ref.~\onlinecite{Bak82}): the incommensurate structure
tends to lock onto rational fractions of the period of the potential.
Instead of a continuous CI transitions one gets a series of
commensurate-commensurate transitions between different lock-in states.

The  interest of Dzyaloshinskii's   model for CI transitions is that it
is driven by a changing strength of the potential alone,  with both
built-in periods remaining constant. The experimentalist has a
convenient handle on the potential term that he can vary by simply
adjusting the external field. Since the ``built-in'' periods do not
change, the transition is continuous with no ``devil's staircase''
behavior.

The other advantage of  the present realization of CI transitions lies
in the fact that the potential energy has a pure sinusoidal form as a
function of the angle $\theta(x)$. The model can be exactly solved and a
quantitative comparison of theory and experiment are possible  for the
entire phase diagram. In most other CI systems one knows only that the
potential energy is a periodic function with a given period, determined
by  the underlying lattice, while  its exact functional form remains
undetermined. Quantitative comparison of experiment  and theory in this
case is restricted to the narrow critical region close to the transition
point.

It is important to note here that while field-induced CI transitions
have been previously observed in a number of magnetic insulators with
incommensurate structure\cite{Heller94,Zhitomirsky95,Knop83,Lebech89}
for various reasons none of them is well described by Dzyaloshinskii's
model. In FeGe (Ref.~\onlinecite{Lebech89}) the problem is that in the
high-symmetry cubic structure applying even a very small  magnetic field
rearranges the spins  so that the spin rotation plane is perpendicular
to the field direction, in violation of Dzyaloshinskii's requirement.
Instead of a 2-nd order Dzyaloshinskii tansition one gets a 1st-order
spin-flip transition from the incommensurate phase directly into the
papramagnetic state. Although this does not happen in compounds like
RbMnBr$_3$ \cite{Heller94,Zhitomirsky95} and CsFeCl$_3$ \cite{Knop83},
the phase behavior there is seriously complicated by the quantum effects
and frustration in the triangular spin lattice.\cite{Jacobs}

In the following  paragraphs we shall demonstrate that the experimental
data on the CI transition in Ba$_{2}$CuGe$_{2}$O$_{7}$ agrees with our
solution to Dzyaloshinskii's model at the quantitative level.
Ba$_{2}$CuGe$_{2}$O$_{7}$ thus appears to be the first system that
exhibits a Dzyaloshinskii-type transition in its original form.

\subsection{Estimates for the critical field}
The appeal of the theory presented above is that it allows for an exact
solution. Its major limitation of course is that it is based on a
classical, rather than quantum spin model. Nevertheless, our theoretical
predictions seem to be in excellent agreement with what is
experimentally observed in Ba$_{2}$CuGe$_{2}$O$_{7}$, even at the
quantitative level. To begin with, the model gives the correct value for
the critical field $H_{c}$. The in-plane exchange parameter $J=0.48$~meV
was previously determined by measuring the spin-wave dispersion
spectrum,\cite{Zheludev96-BACUGEO}. The exchange energy per bond is then
$\tilde{J}=2JS^{2}\approx0.24$~meV. The $g$-values were measured in ESR
experiments: $g_{a}=2.044$ and $g_{c}= 2.474$.\cite{Sasago-ESR} By
substituting $\zeta(0)=0.027$ into Eq.~(\ref{estimate}) we immediately
obtain $H_{c}=3.3$~T, that should be compared to the experimental value
$H_{c}\approx 2.1$~T. Considering that for the theoretical estimate we
used predictions for $T=0$ and ignored quantum effects, a 30\%
consistency is quite acceptable.

\subsection{The soliton lattice}

The most intriguing prediction of our theory for CI transition in
Ba$_{2}$CuGe$_{2}$O$_{7}$ is that the spin structure close to the phase
transition is no longer an ideal spiral, but rather should be viewed as
a lattice of solitons, i.e., domain walls separating regions of
N\'{e}el-like spin arrangement. The soliton lattice is a distinguishing
characteristic of all CI systems. Several comprehensive reviews on the
subject exist, among them papers by Bak\cite{Bak82} and Pokrovsky et
al.\cite{Pokrovsky86} The basic physical mechanism is quite simple. When
the potential term is sufficiently large, but still smaller than the
critical value, it is favorable for the system to have large
``commensurate'' regions. The elastic energy from the ``incommensurate''
term in the Hamiltonian  is partially released by forming domain walls
or phase slips that separate ``commensurate'' domains.

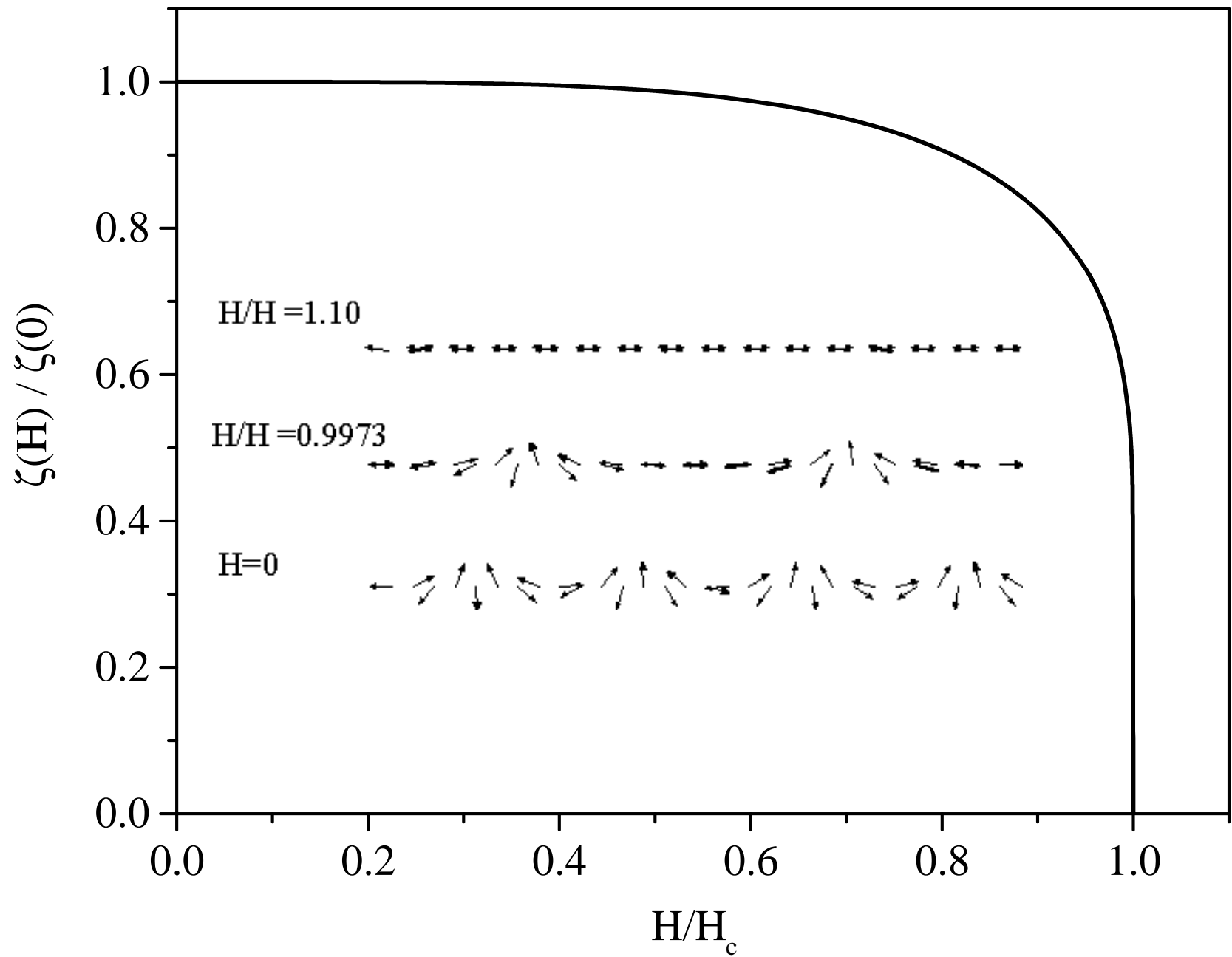
\begin{figure}
\caption{Theoretical field-dependence of the magnetic propagation vector
$\zeta$,
obtained from an exact solution of a classical 1-dimensional
antiferromagnet with Dzyaloshinskii-Moriya interactions in the continuous
limit. Insert: schematic representation of the spiral phase ($H=0$),
soliton lattice ($0<H<H_{c}$) and the spin-flop commensurate state
($H>H_{c}$).}
\label{theory}
\end{figure}

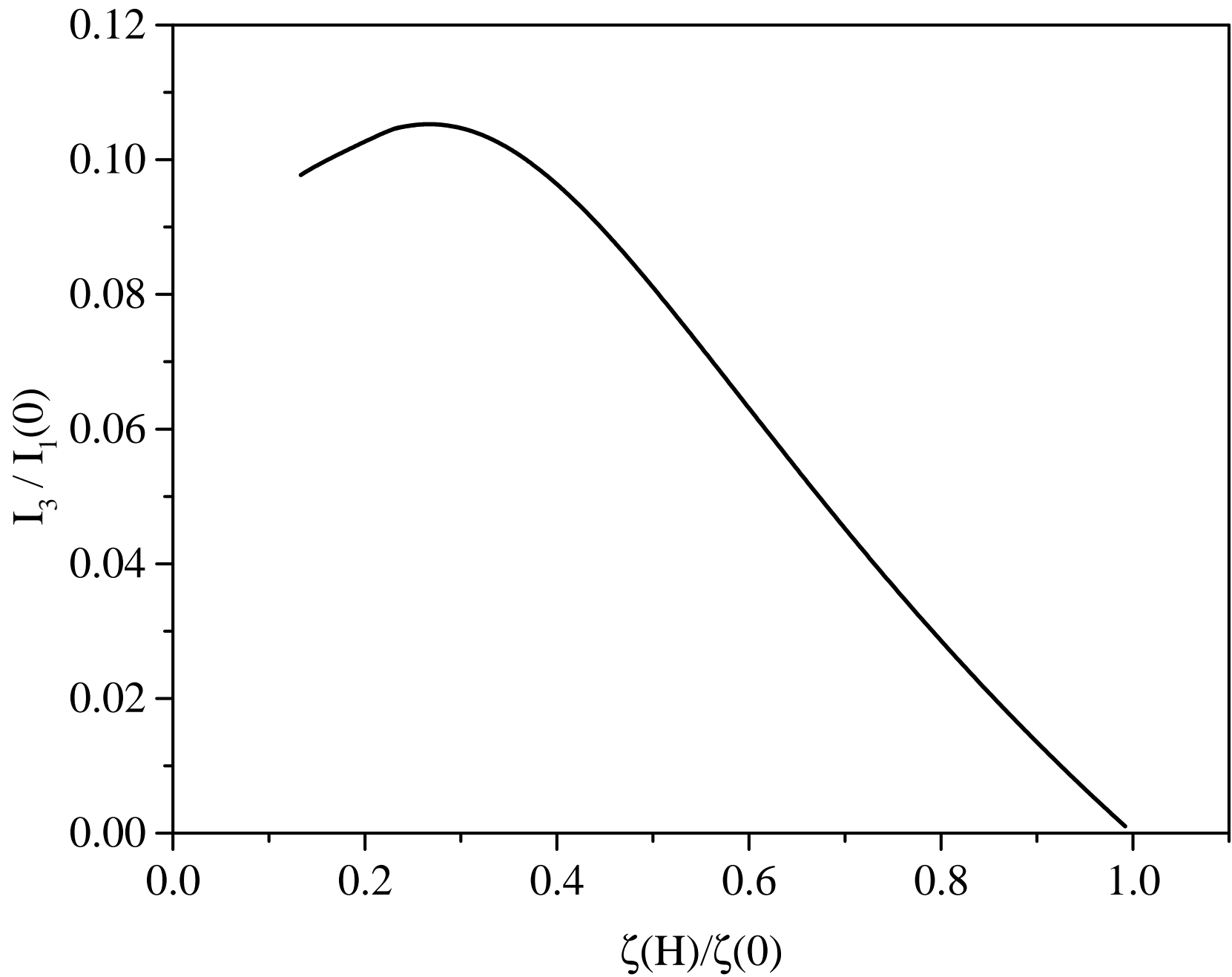
\begin{figure}
\caption{
Theoretically predicted Bragg intensity of the 3rd Fourier
harmonic
of the soliton lattice plotted as a function of $\zeta(H)/\zeta(0)$.}
\label{harmonic}
\end{figure}

From the experimental point of view the soliton lattice concept has
three important consequences. The first is that the incommensurability
parameter $\zeta$ is field-dependent, and $\zeta\rightarrow 0$ {\it
continuously} as $H\rightarrow H_{c}$ (Fig.~\ref{theory}). Moreover, the
transition itself is very unusual: continuous, yet logarithmically steep
at $H_{c}$. The experimental $\zeta(H)$ data collected for the
well-aligned domain (Fig.~\ref{good}, open circles) can be nicely fit by
the theoretical curve shown in Fig.~\ref{theory}, treating $\zeta(0)$
and $H_{c}$ as adjustable parameters. A very good agreement is obtained
with $\zeta(0)=0.027(1)$ and $H_{c}=2.13(2)$~T at $T=2.4$~K (solid line
in Fig.~\ref{good}). All the way up to $\zeta(H)/\zeta(0)\approx 2/3$
the theoretical curve follows the experimetal points closely. In the
very proximity of $H_{c}$ however, a deviation is apparent. This is to
be expected: the soliton lattice becomes infinitely soft at the
transition point and pinning to structural defects leads to a saturation
of $\zeta(H)$ at $H\rightarrow H_{c}$ and effective broadening of
magnetic Bragg reflections (Fig.~\ref{exdata2}). The almost-first-order
transition is very fragile. A small misalignment of the field produces
what we see in the misaligned domain as a first-order transition with a
mixed-phase region and a substantially different form of $\zeta(H)$ in
the vicinity of $H_{c}$.

Inherently related to the unusual logarithmic phase transition is an
anomaly in the magnetic susceptibility at $H_{c}$. We have used the
theoretical expression (\ref{chi}) to fit the experimental $\chi(H)$
shown in Fig.~\ref{mag}. Since we  cannot explicitly take into account
the quantum and thermal spin fluctuations, we have treated $\chi_{||}$,
$\chi_{\perp}$, and $H_c$ as independent fitting parameters. In
addition, we allowed for a linear term $vH$ in $\chi(H)$ that is present
in both the ${\bf H}||{\bf a}$ and ${\bf H}||{\bf c}$ data. This term
empirically accounts for intrinsic non-linearities in magnetization
curves in the quantum (quasi) 2-dimensional AF Heisenber model (Ref.
\onlinecite{Fabricius92}, esp. Fig. 5), that in our case effectively modify
the local susceptibilities $\chi_{\bot}$ and $\chi_{\|}$ at high fields.
The fit is shown in a solid line in Fig.~\ref{mag}. The values of the
fitting parameters are: $\chi_{||}=0.89 \times 10^{-5}$~emu/g,
$\chi_{\perp}=3.43 \times 10^{-5}$~emu/g, $v=1.72 \times
10^{-10}$~emu/g, and $H_c=1.88$~T. This value for $H_{c}$ obtained from
magnetization at $T=2$~K is slightly lower than $H_{c}=2.13(2)$~T
obtained from neutron diffraction at $T=2.4$K.

Finally, an important feature of the magnetic structure at finite fields
is that it no longer is an ideal sinusoidal spiral. As discussed above,
this distortion is characterized by higher-order Bragg harmonics, that
should be observable in neutron diffraction experiments. Although we
have spent some time looking for the 3rd order magnetic satellite during
experiment II, at $T=2.4$~K and for several values of applied field, we
were so far unable to find it. The reason is probably the lack of
intensity. As can be seen from Fig.~\ref{harmonic}, the relative
intensity of the 3rd harmonic is very small except in the immediate
proximity of $H_{c}$, where the 1st satellite itself is broadened and
weakened. We would like to emphasize here that the form of $\zeta(H)$ and
the existence of higher-order harmonics are {\it in essence one and the
same effect}. Since theory agrees so well with experiment as far as
$\zeta(H)$ is concerned, we are confident in that satellites are present
and will be observed in future experimental efforts.

\subsection{System dimensionality}

We finally comment on our measurements of critical exponents. In our
previous work we have shown that there are several hints, including the
temperature dependence of magnetic susceptibility,  to that
Ba$_{2}$CuGe$_{2}$O$_{7}$ should primarily be considered a 2-dimensional
antiferromagnet. This is further confirmed by the measured
order-parameter critical exponent $\beta=0.184$ for the PM$\rightarrow$
commensurate transition, i.e., substantially smaller than in standard
3-D models, where $\beta> 0.3$ for Heisenberg, XY and even Ising
systems. One could naively expect that at high external fields the
critical index would be {\it smaller} than at $H=0$, since the symmetry
of the corresponding Hamiltonian is smaller. Exactly the opposite is
observed experimentally. The only suggestion that we can make at this
point is that Ba$_{2}$CuGe$_{2}$O$_{7}$ falls into a completely
different universality class than conventional magnets.   Considering
the nature of the spiral phase and D-M interactions, we may be dealing
with one of the chiral universality classes.\cite{Kawamura88}  To obtain
further insight into the critical properties of
Ba$_{2}$CuGe$_{2}$O$_{7}$ further field-dependent measurements of the of
$\beta$ and other critical indexes are required.

No matter what the observed critical  exponents are, the long-range
magnetic ordering is still a purely 3-dimensional phenomenon. For a CI
occurring at finite temperature dimensionality is known to play a key
role. In a purely 1D system at non-zero temperature  the ground state
will be destroyed, since the energy required to create a soliton is
finite while the entropy gain proportional to $ln L$ is infinite in
thermodynamic limit. Solitons will be spontaneously created at any
temperature and no sharp transition will occur. In two dimensions, the
effect of thermal  (or quantum) fluctuations is more subtle.  For a
general case Pokrovskii, Talapov and Bak have demonstrated that the
effective soliton-soliton interaction is altered by
fluctuations.\cite{Pokrovsky86} The short range exponential repulsion
$\exp (-\lambda r)$ is replaced by a long range term $1/r^2$. This
modifies the behavior close to the transition point, making the phase
transition a usual second order-type with $\zeta(H) \sim (H_c-H)^{1/2}$.
Only in 3 dimensions the results that we have  derived for the ground
state should  remain valid at finite temperatures. The transition  in
this case is of almost first order, with logarithmic ``corrections''
near the transition point making it continuous.

\subsection{Ideas for future experiments}
Ba$_{2}$CuGe$_{2}$O$_{7}$ is a very interesting system, yet it is
relatively easy to investigate experimentally. The Heisenberg exchange
constants are small, so the magnon dispersion relations could be
measured in the entire Brillouin zone. D-M interactions are relatively
strong and the incommensurability parameter $\zeta(0)$ is sufficiently
large to be easily measurable. $H_{c}\approx 2$~T is also readily
accessible in most types of experiment, even those that require the use
of diffraction-adapted horizontal-field magnets. The work on
Ba$_{2}$CuGe$_{2}$O$_{7}$ is far from being completed. The first
priority is to find the higher-order harmonics of the incommensurate
magnetic peaks and study the temperature dependence of their
intensities. One should also look for satellites around ferromagnetic
zone-centers to see if there is a weak-ferromagnet distortion of the
spiral structure. The effect of magnetic field applied in the $(a,b)$
plane is worth investigating. The critical behavior is not fully
understood and accurate field-dependent measurements of all critical
indexes are highly desirable. Finally, it would be interesting to look
at dynamical propeties of the soliton lattice. Near the critical field
the soliton-soliton interaction is extremely weak, which gives rise to
spin waves with very low velocity.

\section{Conclusion}

We have observed a rare type of CI transition in an antiferromagnetic
insulator with Dzyaloshinskii-Moriya interactions. The transition is
driven exclusively by the changing strength of the commensurate
potential. The latter is directly controlled in an experiment by varying
the magnetic field. A transition of this kind was envisioned over three
decades ago by Dzyaloshinskii, and we have now found that
Ba$_{2}$CuGe$_2$O$_7$ exhibits it in its original form. In addition, we
have extended Dzyaloshinskii's theoretical treatment to derive exact
parametric equations for the field dependence of magnetization and
incommensurability vector.

\acknowledgements
This study was supported in part by NEDO
   (New Energy and Industrial Technology Development Organization)
   International Joint Research Grant and the U.S. -Japan Cooperative
   Program on Neutron Scattering.
Work at Brookhaven National Laboratory was carried out under Contract
   No. DE-AC02-76CH00016,
   Division of Material Science, U.S. Department of Energy.


\end{document}